\documentclass[3p,times,procedia]{elsarticle}
\usepackage{nupha_ecrc}
\usepackage{graphicx}
\usepackage{caption}
\usepackage{subcaption}

\volume{00}

\firstpage{1}

\journalname{Nuclear Physics A}

\runauth{}

\jid{nupha}

\jnltitlelogo{Nuclear Physics A}

\usepackage{amssymb}

 \usepackage{lineno}

\usepackage[figuresright]{rotating}

\begin{document}
	
	\begin{frontmatter}
		
		
		
		\dochead{XXVIIIth International Conference on Ultrarelativistic Nucleus-Nucleus Collisions\\ (Quark Matter 2019)}
		
		\title{Heavy-flavour jet production and correlations with ALICE }
		
		
		\author{Jakub Kvapil, for the ALICE Collaboration}
		
		\address{School of Physics and Astronomy, University of Birmingham, United Kingdom\\Email address: jakub.kvapil@cern.ch}
		\begin{abstract}
			This contribution focuses on the latest studies of heavy-flavour jets performed with the ALICE detector.
			The measurements of the jet momentum fraction carried by the D meson in pp collisions at $\sqrt{s}=5.02$ TeV and $13\ \mathrm{TeV}$ will be presented. The comparison to POWHEG+PYTHIA will be shown. Moreover, the first measurement at the LHC of the jet-momentum fraction carried by the $\mathrm{\Lambda_c^+}$ baryon in pp collisions at $\sqrt{s}=13\ \mathrm{TeV}$ will be reported.
			Production of heavy-flavour jets in pp collisions at $\sqrt{s}=5.02\ \mathrm{TeV}$ will be also addressed with charged jets tagged by heavy-flavour hadron decay electrons and beauty-hadron decay vertices.
			The nuclear modification factor of heavy-flavour tagged jets in p–Pb collisions at $\sqrt{s_{\mathrm{NN}}}=5.02\ \mathrm{TeV}$ will be presented.
		\end{abstract}
		
		\begin{keyword}
			Heavy-flavour, jets, pQCD, nuclear modification, parallel jet momentum
			
		\end{keyword}
		
	\end{frontmatter}
	
	
	\section{Introduction}
	\label{}
	Heavy quarks (charm and beauty) are mostly produced in hard partonic scattering processes in the early stages of the collisions. Because of their large mass the production cross section can be calculated using perturbative Quantum Chromodynamics (pQCD) down to $p_{\mathrm{T}}=0$. Therefore, they are an ideal probe of the properties of Quark-Gluon Plasma (QGP) created in heavy-ion collisions. Measurements of heavy-flavour tagged jets give direct access to the parton kinematics and can provide information on heavy-quark energy loss in the QGP, in particular on how the radiated energy is dissipated in the medium. Studies of angular correlations of heavy-flavour particles with charged particles allow us to characterise the heavy-quark fragmentation process and its possible modification in a hot nuclear-matter environment.
	
	Measurements in pp collisions provide constraints on pQCD-based models and are necessary as a reference for the interpretation of heavy-ion collision results.
	Studies in p--Pb collisions can give insight on how heavy-quark production and hadronisation into jets are affected by cold nuclear-matter effects.

	\section{Results}
	\subsection{$\mathrm{D}$-meson and $\Lambda_c^{+}$-baryon tagged jets}
\begin{figure*}[t!]
		\centering
		\begin{subfigure}[t]{0.45\textwidth}
			\centering
			\includegraphics[height=1.9in]{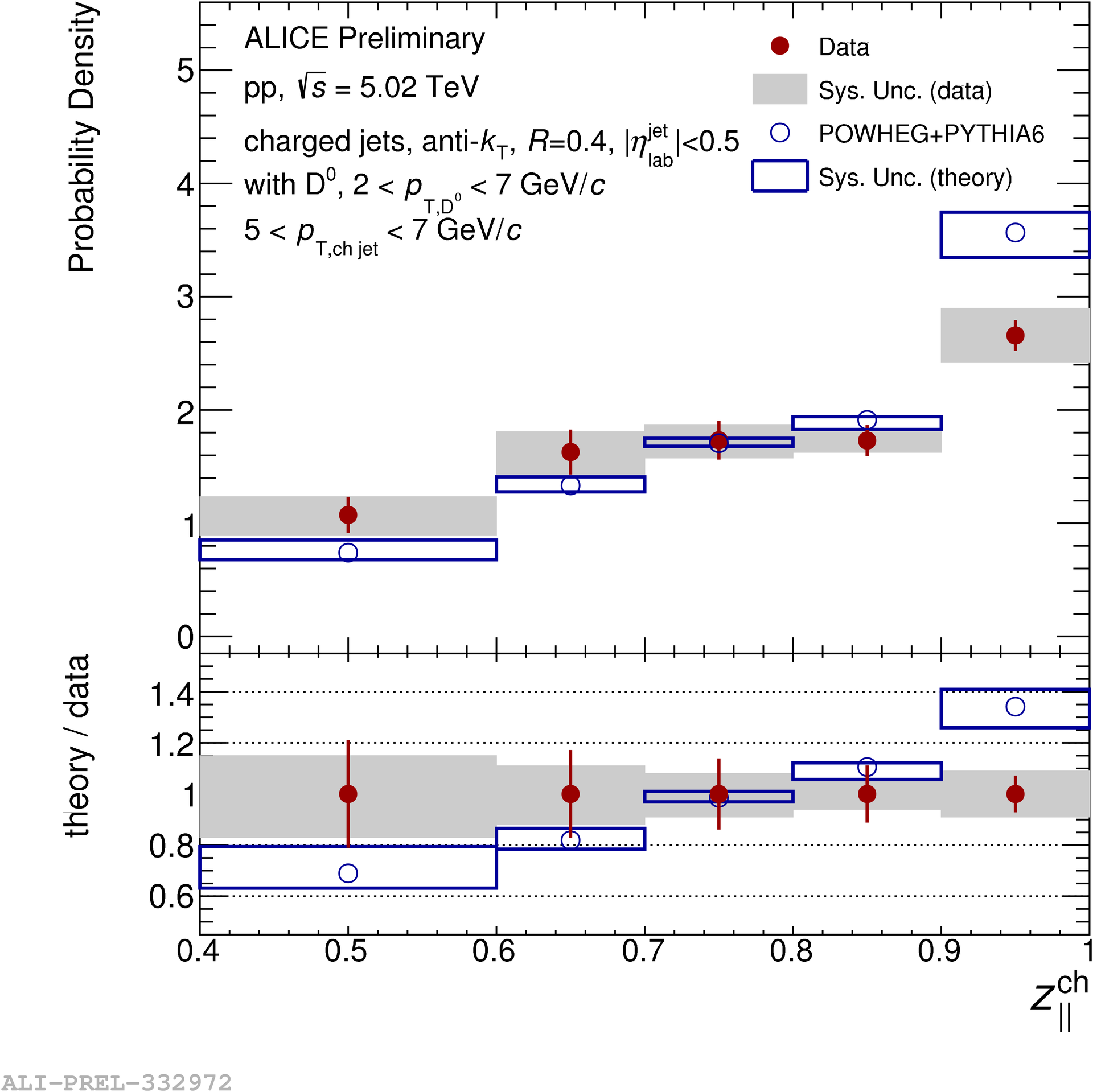}
			\caption{}
		\end{subfigure}%
		\begin{subfigure}[t]{0.45\textwidth}
			\centering
			\includegraphics[height=1.9in]{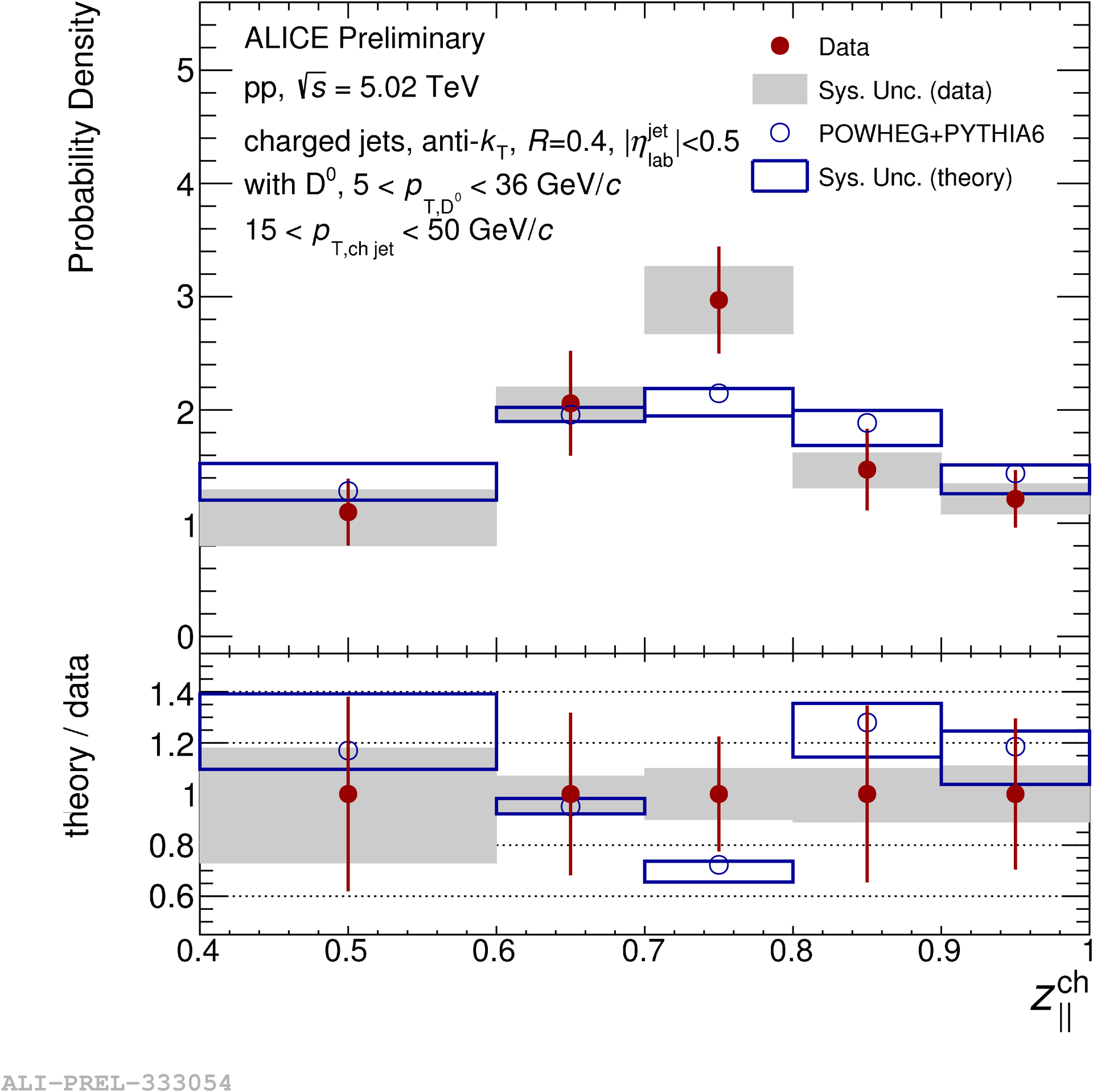}
			\caption{}
		\end{subfigure}
		\caption{Probability density distribution of the jet momentum fraction, $z^{ch}_{\parallel}$, carried by $\mathrm{D^0}$ mesons measured in pp collisions at $\sqrt{s}=5.02\ \mathrm{TeV}$ for (a) $5<p_{\mathrm{T,jet}}<7\ \mathrm{GeV/}c$  and (b) $15<p_{\mathrm{T,jet}}<50\ \mathrm{GeV/}c$, compared to POWHEG+PYTHIA6 predictions \cite{Frixione:2007vw,Sjostrand:2006za}.}
		\label{fig:D}
	\end{figure*}

\begin{figure*}[t!]
		\centering
		
		\begin{subfigure}[t]{0.45\textwidth}
			\centering
			\includegraphics[height=1.80in]{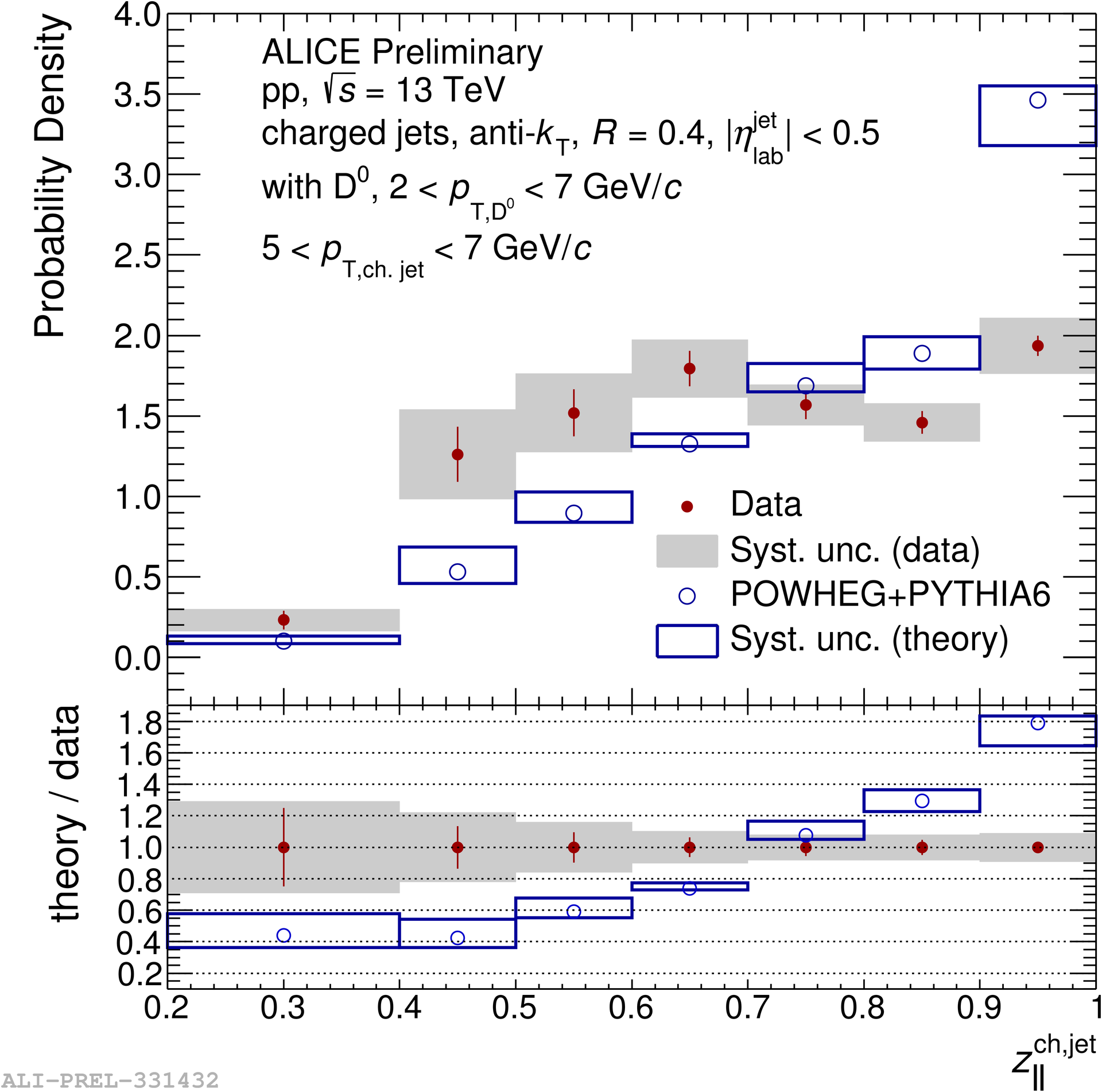}
			\caption{}
		\end{subfigure}
		\begin{subfigure}[t]{0.45\textwidth}
			\centering
			\includegraphics[height=1.80in]{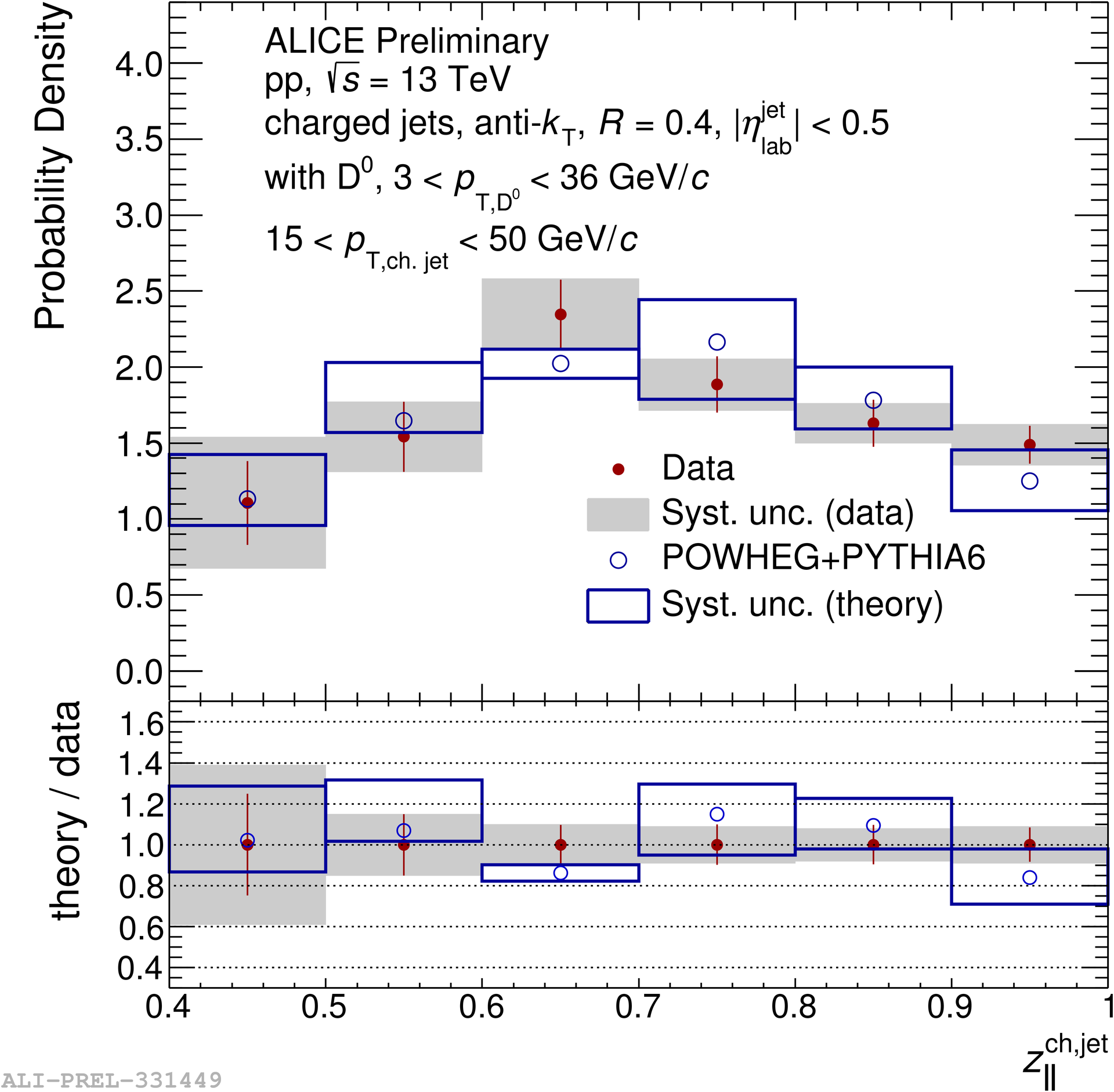}
			\caption{}
		\end{subfigure}
		\caption{Probability density distribution of the jet momentum fraction, $z^{ch}_{\parallel}$, carried by $\mathrm{D^0}$ mesons measured in pp collisions at $\sqrt{s}=13\ \mathrm{TeV}$ for (a) $5<p_{\mathrm{T,jet}}<7\ \mathrm{GeV/}c$  and (b) $15<p_{\mathrm{T,jet}}<50\ \mathrm{GeV/}c$, compared to POWHEG+PYTHIA6 predictions \cite{Frixione:2007vw,Sjostrand:2006za}.}
		\label{fig:D2}
	\end{figure*}

	Jets are reconstructed using the anti-$k_\mathrm{T}$ \cite{Cacciari:2011ma} algorithm and are tagged as $\mathrm{D^0}$-meson or $\mathrm{\Lambda_c^{+}}$-baryon jets if they contain a fully reconstructed $\mathrm{D^0}$ meson or $\mathrm{\Lambda_c^{+}}$ baryon. The studied decay chanels are $\mathrm{D^0}\rightarrow \mathrm{K^-}\mathrm{\pi^+}$ (B.R. 3.89\%) and $\mathrm{\Lambda_c^{+}}\rightarrow \mathrm{p}{\rm K}_{\rm S}^{0}$ (B.R. 1.59\%) with corresponding charge conjugates. The sideband method is used for subtracting the combinatorial background and extracting the heavy-flavour-jet raw signals. The feed-down contribution from beauty hadron decays is estimated and subtracted using a simulation performed with POWHEG+PYTHIA6 \cite{Frixione:2007vw,Sjostrand:2006za}, as detailed in \cite{Acharya:2019zup}. A Bayesian unfolding procedure was performed using the RooUnfold package \cite{Adye:2011gm} to correct for detector effects affecting the reconstructed jet momentum. The parallel jet momentum fraction, $z^{ch}_{\parallel}=\frac{\vec{p}_\mathrm{D}\cdot \vec{p}_{\mathrm{ch. jet}}}{\vec{p}_\mathrm{ch. jet}\cdot \vec{p}_\mathrm{ch. jet}}$, is proportional to the emitted angle of the $\mathrm{D^0}$ meson with respect to the jet axis. The $z^{ch}_{\parallel}$ probability density of $\mathrm{D^0}$-tagged jets with $5<p_{\mathrm{T,jet}}<7\ \mathrm{GeV/}c$ and $15<p_{\mathrm{T,jet}}<50\ \mathrm{GeV/}c$ is reported in Fig. \ref{fig:D} and \ref{fig:D2} for pp collisions at $\sqrt{s}=5.02\ \mathrm{TeV}$ and $\sqrt{s}=13\ \mathrm{TeV}$, respectively. The data are well described by POWHEG+PYTHIA6 at large $p_{\mathrm{T,jet}}$, whilst at lower momentum the prediction shows a harder fragmentation than the data. In Fig. \ref{bje} (a) the $z^{ch}_{\parallel}$ probability density of $\mathrm{\Lambda^+_c}$-tagged jets is compared to expectations from POWHEG+PYTHIA6 \cite{Frixione:2007vw,Sjostrand:2006za}, PYTHIA8 \cite{Sjostrand:2007gs}, and PYTHIA8 with string formation beyond leading-colour approximation \cite{Christiansen:2015yqa}. The latter model expects a softer $z^{ch}_{\parallel}$ distribution than the others, in better agreement with the data.

	\subsection{b-jets}
\begin{figure*}[t!]
		\centering
				\begin{subfigure}[t]{0.5\textwidth}
			\centering
			\includegraphics[height=1.8in]{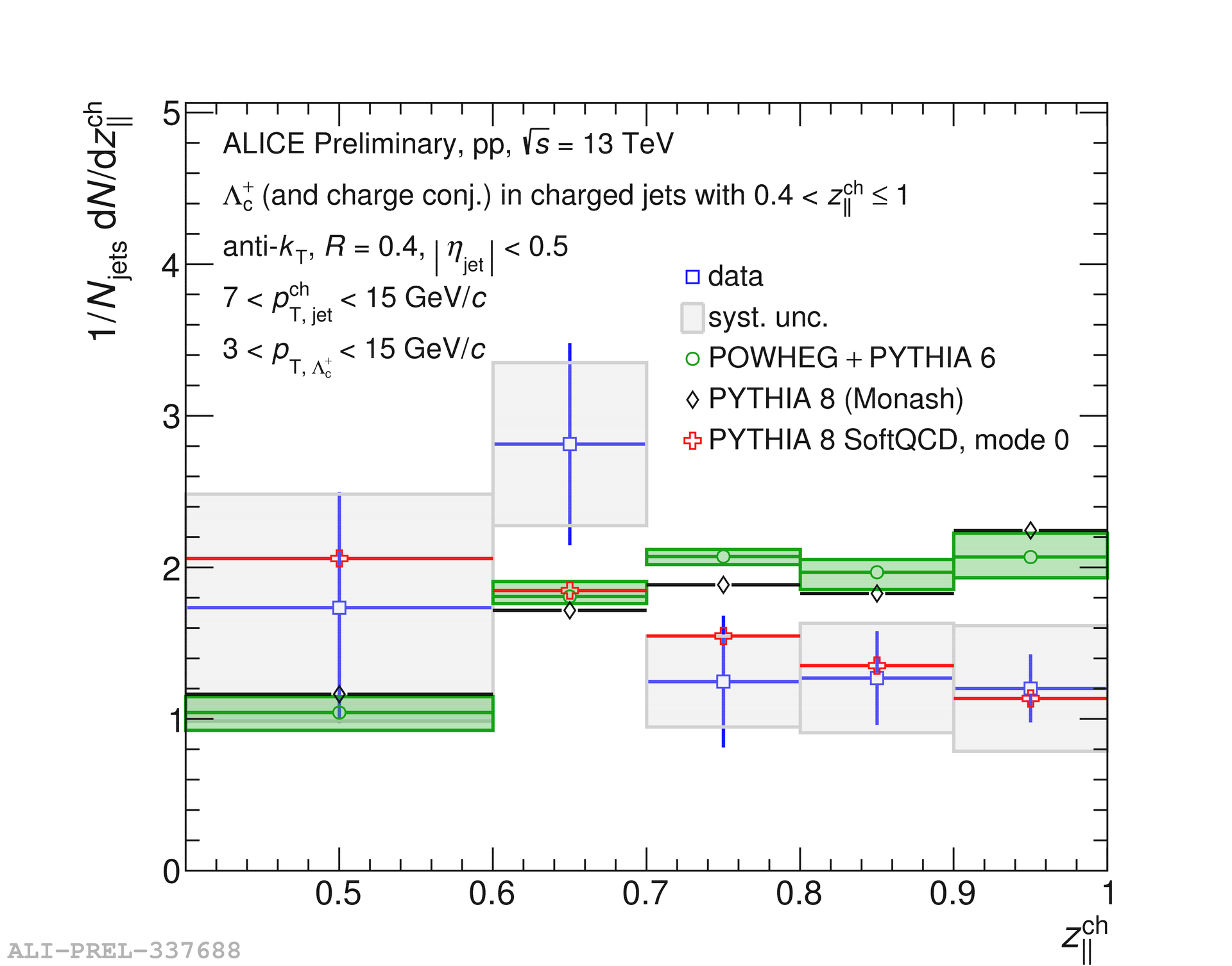}
			\caption{}
		\end{subfigure}%
		\begin{subfigure}[t]{0.5\textwidth}
			\centering
			\includegraphics[height=1.8in]{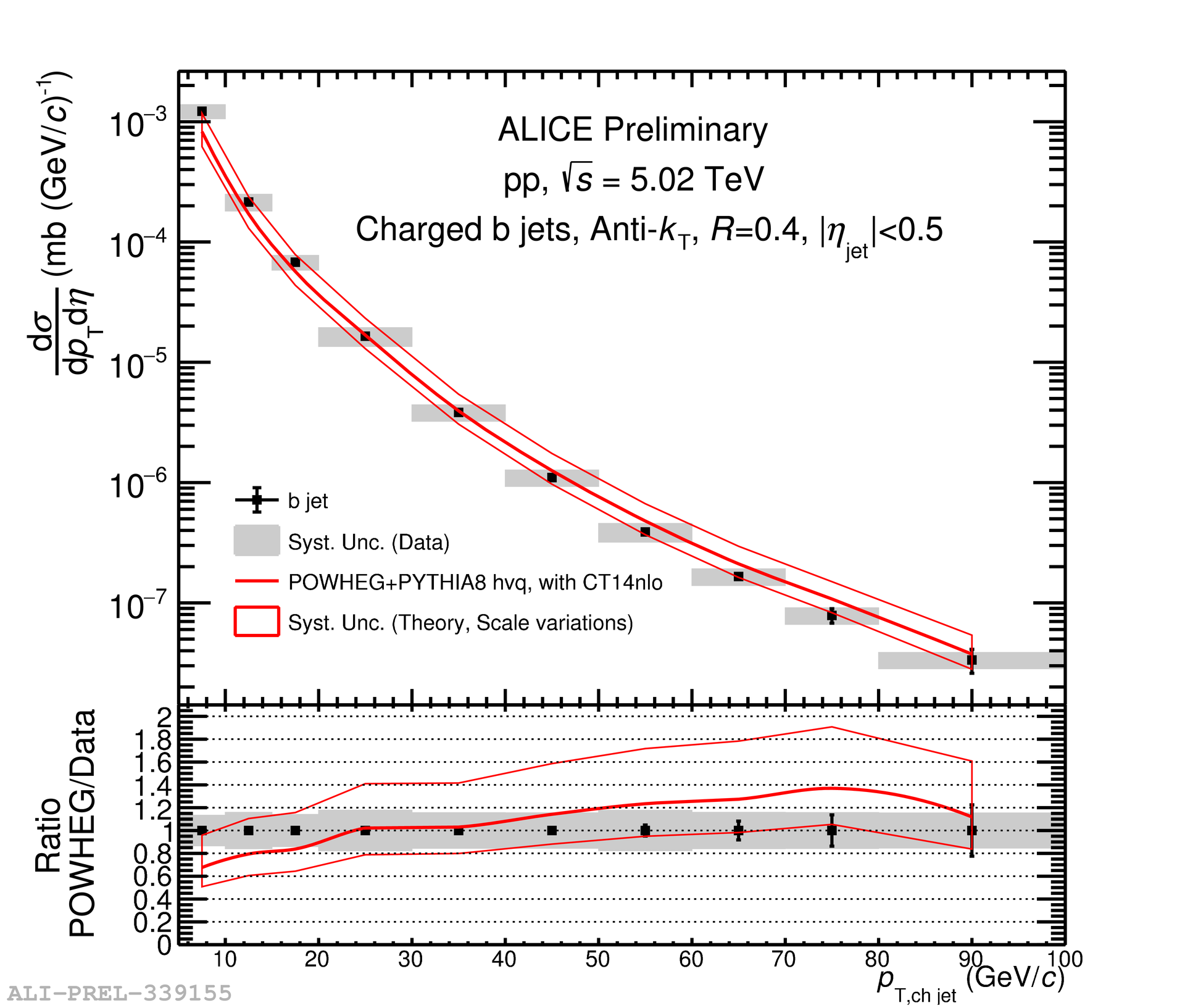}
			\caption{}
		\end{subfigure}
		\caption{(a) Probability density distribution of the jet momentum fraction, $z^{ch}_{\parallel}$, carried by the $\mathrm{\Lambda_c^+}$ baryons measured in pp collisions at $\sqrt{s}=13\ \mathrm{TeV}$ for $7<p_{\mathrm{T,jet}}<15\ \mathrm{GeV/}c$, compared to POWHEG+PYTHIA6 \cite{Frixione:2007vw,Sjostrand:2006za} and PYTHIA8 \cite{Sjostrand:2007gs,Christiansen:2015yqa} predictions. (b) Comparison of the $p_{\mathrm{T}}$-differential cross section of b-jets in pp collisions at $\sqrt{s}=5.02\ \mathrm{TeV}$ with expectations from POWHEG+PYTHIA8 \cite{Frixione:2007vw,Sjostrand:2007gs}.}
		\label{bje}
	\end{figure*}

\begin{figure*}[t!]
		\centering
		\begin{subfigure}[t]{0.5\textwidth}
			\centering
			\includegraphics[height=1.8in]{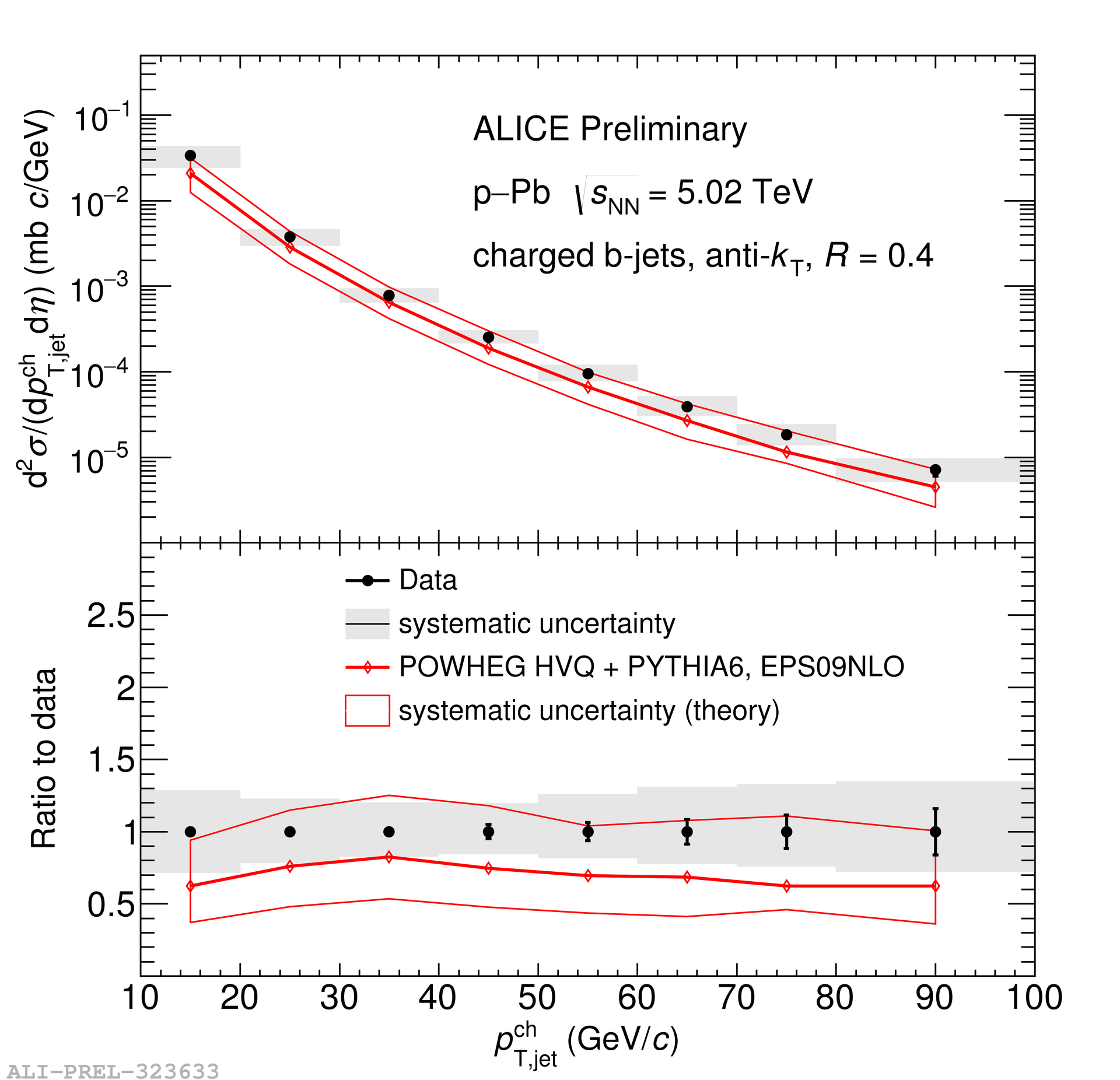}
			\caption{}
		\end{subfigure}%
		\begin{subfigure}[t]{0.5\textwidth}
			\centering
			\includegraphics[height=1.82in]{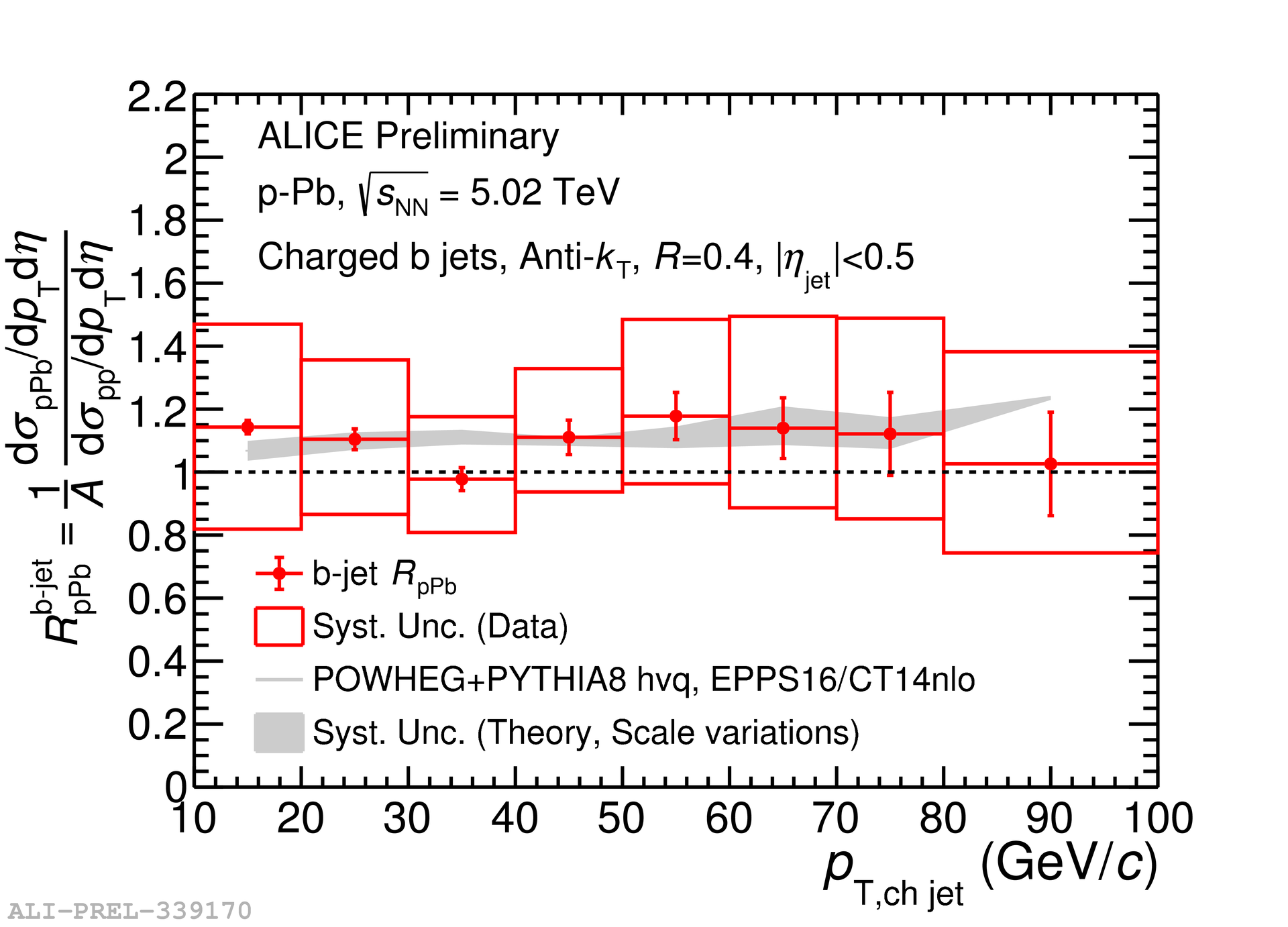}
			\caption{}
		\end{subfigure}
		\caption{(a) Comparison of the b-jet $p_{\mathrm{T}}$-differential cross section in p--Pb collisions at $\sqrt{s_{\mathrm{NN}}}=5.02\ \mathrm{TeV}$, compared to POWHEG+PYTHIA6 \cite{Frixione:2007vw,Sjostrand:2006za} predictions with EPS09NLO nuclear PDF \cite{Eskola:2009uj}. (b) Nuclear modification factor, $R_{\mathrm{pPb}}$, of b-jets at $\sqrt{s_{\mathrm{NN}}}=5.02\ \mathrm{TeV}$, compared to POWHEG+PYTHIA8 \cite{Frixione:2007vw,Sjostrand:2007gs} predictions with EPSS16 nuclear PDF \cite{Eskola:2016oht}.}
		\label{fig:L}
	\end{figure*}

	Two methods are used to identify b-jets, both exploiting the large distance between the b-hadron decay point (secondary vertex) and the collision point (primary vertex). In the first, used in the analysis of pp collisions, tracks inside jets are sorted by their impact parameter to the primary vertex and the jet is tagged as a b-jet if the $\mathrm{2^{nd}}$ most displaced track exceeds a given threshold. In p--Pb collisions 3-prong vertices are reconstructed with the jet tracks and b-jets are identified by the presence of a vertex compatible with a displaced decay topology. The $p_{\mathrm{T}}$-differential cross section of b-jets is shown in pp collisions at $\sqrt{s}=5.02\ \mathrm{TeV}$ in Fig. \ref{bje} (b) and in p--Pb collisions at $\sqrt{s_{\mathrm{NN}}}=5.02\ \mathrm{TeV}$ in Fig. \ref{fig:L} (a). The measurements are well described in both collision systems by POWHEG+PYTHIA predictions that include the usage of nuclear PDF \cite{Eskola:2009uj} for the p-Pb case. The nuclear modification factor of b-jets, defined as the ratio of cross sections in p--Pb and pp collisions, the latter scaled by lead nucleon number, is shown in Fig. \ref{fig:L} (b). The b-jet production appears not to be affected by cold nuclear matter effects within current uncertainties.

	\subsection{Heavy-Flavour-decay-electron tagged jets}
	
\begin{figure*}[t!]
		\centering
		
		\begin{subfigure}[t]{0.48\textwidth}
			\centering
			\includegraphics[height=1.8in]{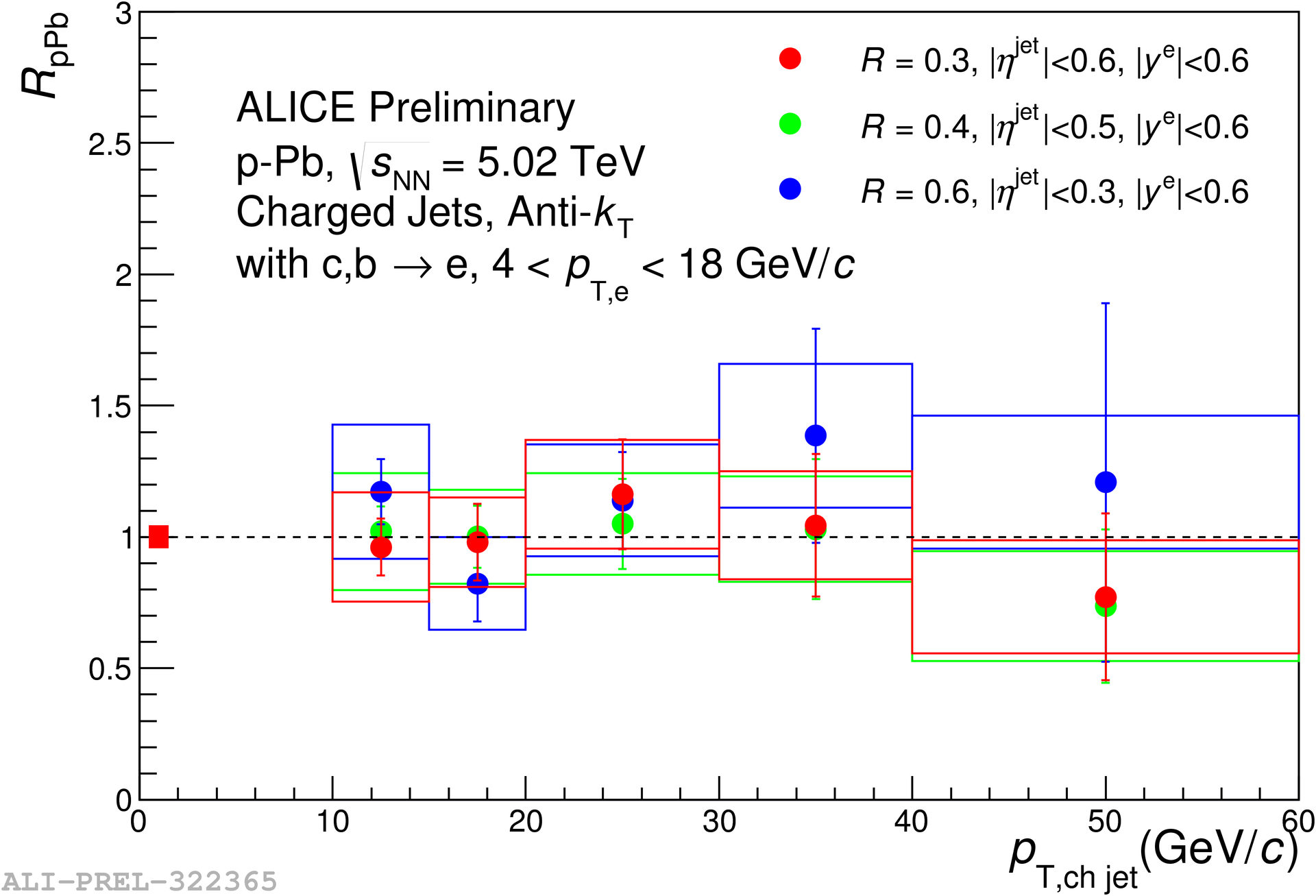}
			\caption{}
		\end{subfigure}
		\begin{subfigure}[t]{0.48\textwidth}
			\centering
			\includegraphics[height=1.8in]{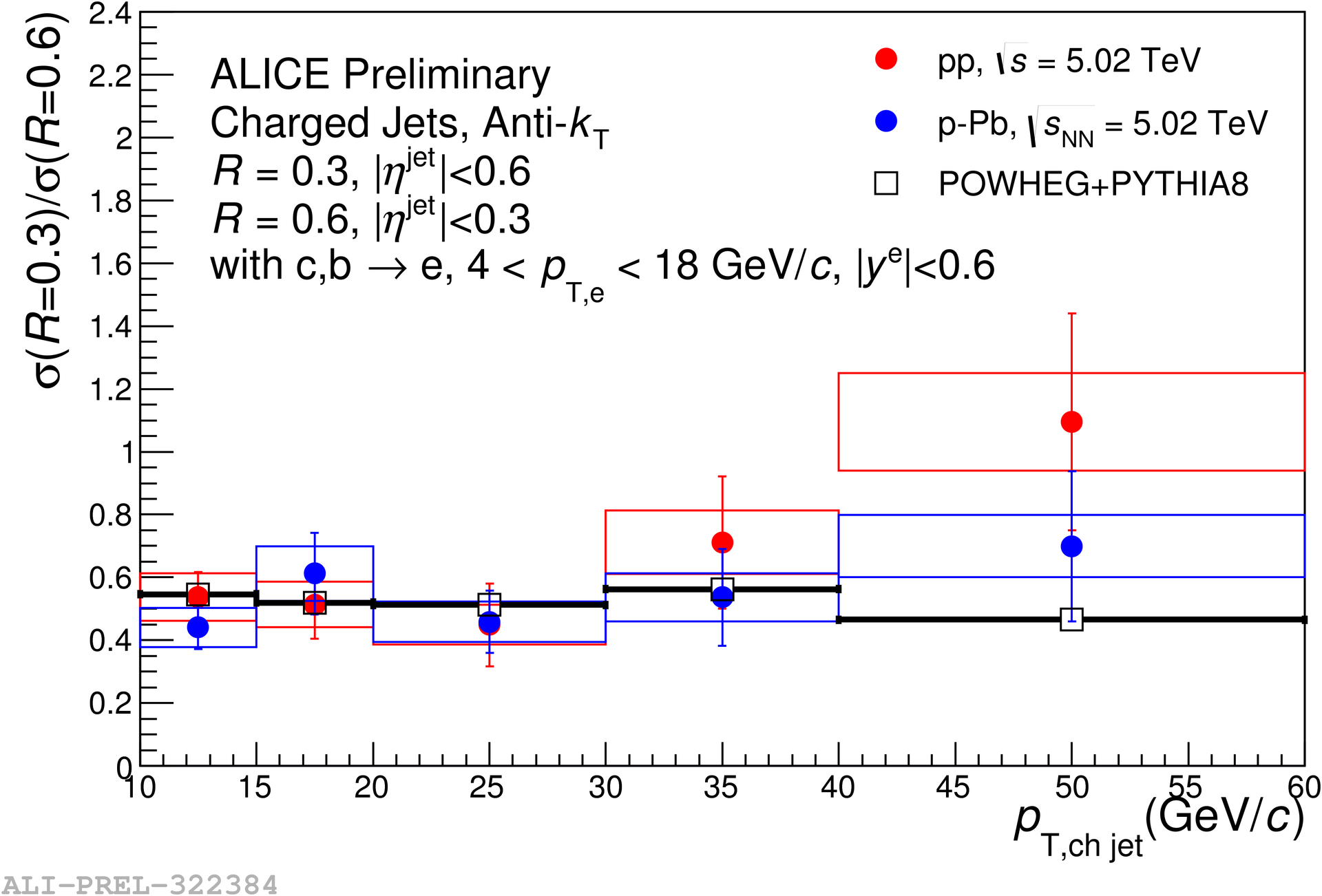}
			\caption{}
		\end{subfigure}
		\caption{(a) Nuclear modification factor, $R_{\mathrm{pPb}}$, of heavy-flavour electron jets at $\sqrt{s_{\mathrm{NN}}}=5.02\ \mathrm{TeV}$ for jet resolution parameters $R=0.3$, $0.4$, $0.6$. (b) Ratio of cross section of jets containing electrons from heavy-flavour hadron decays with resolution parameter $R=0.3$ and $R=0.6$ for pp and p--Pb collisions at $\sqrt{s_{\mathrm{NN}}}=5.02\ \mathrm{TeV}$.}
		\label{hfe}
	\end{figure*}

	In this analysis, heavy-flavour jets are tagged by requiring an electron originating from a heavy-flavour hadron decay among the jet constituents. Previous measurements of heavy-flavour decay electrons in small systems showed a positive elliptic flow $v_2$ \cite{Acharya:2018dxy}, that could be induced by final-state effects. These effects may affect also the jet $p_{\mathrm{T}}$ spectrum, introducing modifications possibly dependent on the jet resolution parameter $R$. The nuclear modification factors of heavy-flavour decay electron jets, shown in Fig. \ref{hfe} (a) for $R=0.3$, $R=0.4$ and $R=0.6$ do not exhibit deviations from unity. The ratio of the $p_\mathrm{T}$-differential cross sections for $R=0.3$ and $R=0.6$ for pp and p--Pb collisions are consistent within uncertainties. The data do not support the presence of large final-state effects on heavy-flavour decay electron jets in p--Pb collisions.

	\subsection{D-meson-hadron correlation}
	The angular correlation of $\mathrm{D^0}$, $\mathrm{D^+}$ and $\mathrm{D^{*+}}$ mesons with charged particles were studied. The correlation functions and the near- and away-side peak properties are found to be consistent in pp and p--Pb collisions, showing no modifications due to nuclear effects within uncertainties \cite{Acharya:2019icl}. 
		
	\section{Summary}
	New measurements of the parallel jet momentum fraction  of $\mathrm{D^0}$-tagged jets in pp collisions at $\sqrt{s}=13\ \mathrm{TeV}$ and $\sqrt{s}=5.02\ \mathrm{TeV}$ were presented. They hint a softer fragmentation at low $p_\mathrm{T,jet}$ than what expected from POWHEG+PY\-THIA. A similar trend is deduced from the first measurement at the LHC of $\mathrm{\Lambda_c^+}$-tagged jets. 
	As the $R_\mathrm{pPb}$ of b-jets in p--Pb collisions at $\sqrt{s_{\mathrm{NN}}}=5.02\ \mathrm{TeV}$ indicate, the production of b-jets does not appear to be affected by cold nuclear matter effects within current uncertainties.
	Furthermore, the $R_\mathrm{pPb}$ of heavy-flavour decay electron tagged jets shows no dependency on the jet resolution parameter $R$.

	
	
	
	
	\bibliographystyle{elsarticle-num}
	\bibliography{CiteDatabase}
	
	
	
	
	
	
	
\end{document}